\documentclass[twocolumn]{aastex631}

\usepackage{graphicx}	% Including figure files
\usepackage{amsmath}	% Advanced maths commands
\usepackage{amssymb}	% Extra maths symbols
\usepackage{xcolor}
\usepackage{multirow}
\usepackage{soul}

\def\bx{\boldsymbol{x}}
\def\kapT{\kappa_{T}}
\def\kapmu{\kappa_{\mu}}

\def\bk{\boldsymbol{k}}
\def\lambdamax{\lambda_{\max}}
\def\lambdad{\lambda_{\rm d}}

\def\kd{k_{\rm d}}

\def\BV{Brunt-V\"ais\"al\"a~}
\def\NT{N_{T} }
\def\Nmu{N_{\mu} }
\def\kT{k_{T} }
\def\knu{k_{\nu} }
\def\kH{k_{H} }
\def\kmax{k_{\max} }

\def\Eq{Equation}
\def\Eqs{Equations}
\def\beq{\begin{equation}}
\def\eeq{\end{equation}}

\begin{document}

\title{Unifying Transport Models of Thermohaline Convection in Stars}

\author{Valentin A. Skoutnev}
\affiliation{Physics Department and Columbia Astrophysics Laboratory, Columbia University, 538 West 120th Street New York, NY 10027,USA}



\begin{abstract}

Thermohaline convection is a standard chemical mixing process in stellar interiors, yet its mixing efficiency is not fully settled. Competing theories predict turbulent diffusion coefficients, $D_\mu$, that can differ by orders of magnitude, leading to uncertainties in stellar models and interpretations of observations. This Letter explores a potential resolution to existing discrepancies. We first complete the linear stability theory and identify two types of unstable modes: slow growing modes at large length scales and fast growing modes at small length scales. We then reevaluate $D_\mu$ considering the full spectrum of unstable modes and find that it can self-consistently interpolate between previously proposed theoretical scalings across the instability parameter space. The question of thermohaline mixing efficiency in stars may be settled by future simulations that quantify the scale-dependent contributions of fast and slow modes to $D_\mu$ and determine how the modes dominating the transport change across parameter space.

\end{abstract}

\keywords{ Astrophysical fluid dynamics (101) —  Stellar Physics (1621) — Stellar interiors (1606) }


\section{Introduction}

Thermohaline (fingering) convection is a mixing process  in regions where the mean molecular weight $\mu$ increases toward the surface, but the subadiabatic thermal stratification maintains stability to Ledoux convection (for reviews, see \cite{salaris2017chemical,garaud2018double}). Such ``inverted" $\mu-$gradients can form deep inside stellar interiors by nuclear burning or near surface layers by accretion of metal-rich material. Thermohaline mixing has been invoked to explain surface abundance changes in evolving stars \citep{charbonnel2007thermohaline,cantiello2010thermohaline,charbonnel2010thermohaline,stancliffe2010effects,placco2014carbon,lagarde2019gaia,magrini2021gaia}, dilute surface abundances during accretion onto stars \citep{vauclair2004metallic,stancliffe2007carbon,garaud2011happened,theado2012metal,sevilla2022long,behmard2023planet} and white dwarfs \citep{deal2013accretion,bauer2018increases,bauer2019polluted,wachlin2022new,cresswell20253d}, redistribute composition after stellar mergers \citep{glebbeek2008evolution,schneider2025theory}, and transport magnetic fields in crystallizing white dwarfs \citep{castro2024magnetic,fuentes2024short}. However, its mixing efficiency is still debated.

A major challenge is the extreme parameter regime in stars. Stellar plasma is characterized by  small  compositional and momentum diffusivities, $\kapmu$ and (kinematic viscosity) $\nu$, compared to the thermal diffusivity, $\kapT$. Their ratios are key parameters: $\tau=\kapmu/\kapT$ and (Prandtl number) $Pr=\nu/\kapT$, with typical values of $\tau\lesssim Pr={\cal O}(10^{-6})$ in red giants \citep{garaud2015excitation} and $\tau\ll Pr={\cal O}(10^{-2})$ in white dwarfs interiors \citep{montgomery2024fluid}.  At low $Pr$, heat diffuses much faster than momentum, so compositionally buoyant parcels can effectively ignore the thermal stratification and accelerate to significant vertical speeds with negligible viscous effects. Thermohaline convection can therefore be highly turbulent, restricting simulations to regimes far from stellar parameters. By contrast, terrestrial oceans have $\tau\ll Pr={\cal O}(1)$ and viscous effects control the flow dynamics. 

The main question in the low $Pr$ regime is the scaling and magnitude of the turbulent compositional diffusivity $D_\mu$, which relates the  compositional flux to the local $\mu-$gradient. The traditional prescription for $D_\mu$  developed by \citet{ulrich1972thermohaline} and \cite{kippenhahn1980time} predicts efficient mixing and is widely adopted in 1D stellar evolution models. It can reproduce surface-abundance trends of low mass stars on the red giant branch (RGB) where observational constraints are significant \citep{gratton2000mixing,shetrone2019constraining}, though only with a boosted nondimensional prefactor $C_t={\cal O}(10^3)$ (e.g., \citealt{charbonnel2007thermohaline,lattanzio2015numerical}). Recent numerical and theoretical efforts \citep{traxler2011numerically,brown2013chemical,zemskova2014fingering,fraser2026bridging}, however, have converged to a different scaling for $D_\mu$ following the ``parasitic saturation" model of \citep{brown2013chemical}. It predicts a substantially lower mixing efficiency that cannot account for the observations \citep{wachlin2014fingering}. If thermohaline mixing is the responsible mixing mechanism in RGB stars, this tension suggests either that simulations have not reached appropriately extreme parameter regimes or that missing physics robustly enhances thermohaline mixing in most stars, e.g. rotation \citep{sengupta2018effect} or magnetic fields \citep{harrington2019enhanced,fraser2024magnetized}. Otherwise, alternative mixing processes must be active \citep{schwab2020helium,tayar2022thermohaline,denissenkov2024enhanced,wu2025physical}. The question of mixing efficiency also extends to other astrophysical applications, including later phases of stellar evolution \citep{cantiello2010thermohaline}, binary interactions \citep{renzo2021evolution}, and white dwarfs \citep{bauer2019polluted,montgomery2024fluid}.

This Letter reinvestigates the basic theory of thermohaline convection at low $Pr$. We first revisit the linear theory and identify unstable modes across a wide range of spatial scales (Section~\ref{sec:LinearTheory}). The modes naturally separate into two types: slow, large-scale and fast, small-scale modes. We then examine the turbulent transport provided by both mode types and propose a new model for $D_\mu$ (Section~\ref{sec:TransportModels}). Implications of the results and future directions are then discussed (Section~\ref{sec:Discussion}).

\section{Linear theory}
\label{sec:LinearTheory}
\begin{figure*}
    \centering
    \includegraphics[width=\linewidth]{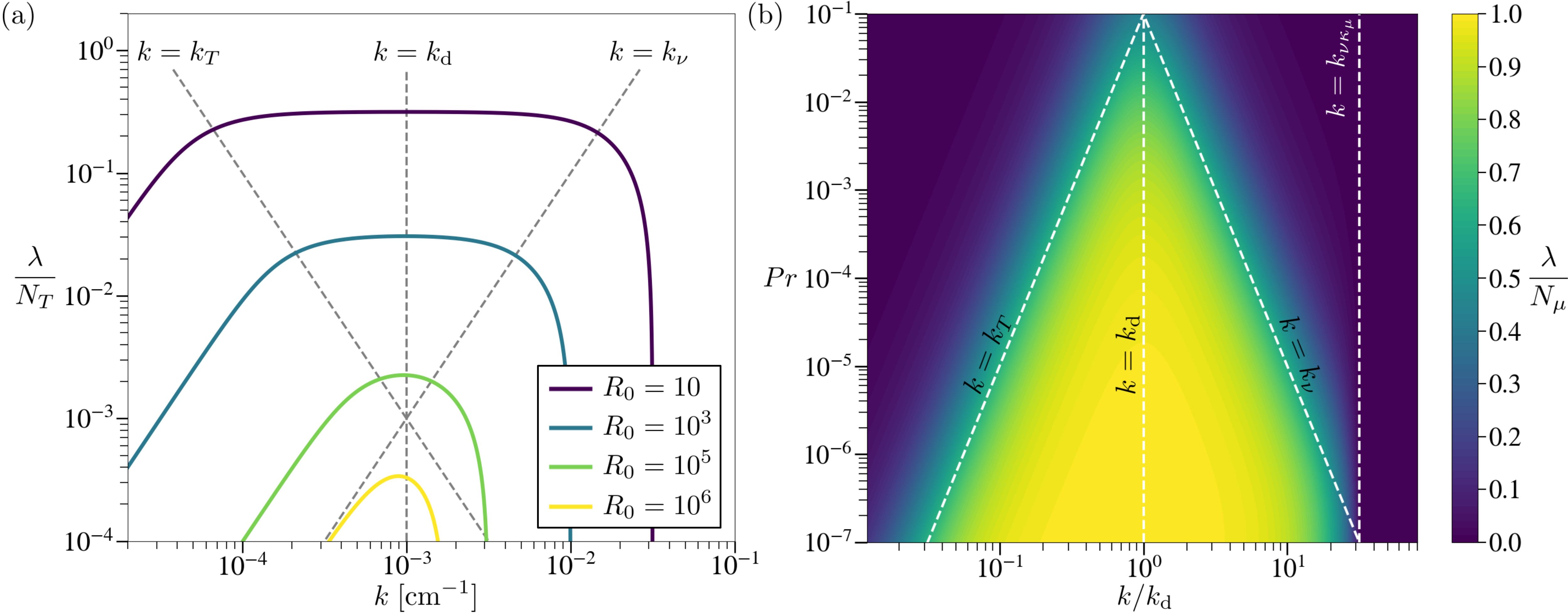}
    \caption{
    Thermohaline growth rate $\lambda$ versus wavenumber $k$ obtained by numerically solving the dispersion relation (\Eq~\ref{eq:D0}) for representative parameters inside a red giant ($\NT=0.1\,\rm s^{-1}$, $\nu=100\,\mathrm{cm}^2\mathrm{s}^{-1}$, $\tau=10^{-7}$). Panel (a) shows $\lambda(k)$ for different density ratios $R_0=\NT^2/\Nmu^2$ at a fixed Prandtl number $Pr=10^{-6}$. For smaller density ratios $R_0<Pr^{-1}/4$, the growth rate plateaus across a range of wavenumbers $\kT<k<\knu$ (the fast modes) with a maximum value $\lambda\approx\NT/R_0^{1/2}=\Nmu$. For larger density ratios $Pr^{-1}/4< R_0<\tau^{-1}$, the plateau disappears and the growth rate maximum is attained only at a single wavenumber $k\approx\kd$ with a reduced value of $\lambda\approx \NT/2R_0Pr^{1/2}$. Panel (b) examines the effect of $Pr$ on the growth rate at fixed $R_0=10$. As $Pr$ decreases,  the plateau widens and thermohaline convection is driven across a broader range of scales.
    }
    \label{fig:GR}
\end{figure*}

Thermohaline convection is characterized by subsonic flow speeds and short vertical length scales relative to the local pressure scale height. Under these conditions, evolution of the velocity, temperature, and composition fields can be modeled using the Boussinesq approximation \citep{spiegel1960boussinesq}. The background stratification is then characterized by two \BV frequencies that we define to be positive and real: $\NT>0$ for the stable temperature gradient and $\Nmu>0$ for the unstable compositional gradient\footnote{Our definition of $\Nmu^2=g\varphi\partial_z\ln\mu$ differs from the standard convention \citep{kippenhahn1990stellar} by a minus sign, where $g$ is the gravitational acceleration and $\varphi=(\partial \ln\rho/\partial\ln\mu)|_{P,T}$.}. Thermohaline convection can occur when the background satisfies the Ledoux criterion $\NT^2-\Nmu^2>0$ for stability against overturning convection.

The linear stability analysis proceeds by obtaining the dispersion relation from the linearized dynamical equations. The analysis in a triply periodic Cartesian domain using normal modes of the form  $\propto \exp(\lambda t+i\bk\cdot \bx)$ is detailed in previous works \citep{stern1960salt,baines1969thermohaline,radko2013double}. The most unstable modes have horizontal wavevectors ($\bk\perp\hat z$, where gravity is along $-\hat z$) corresponding to vertically rising/falling fluid motions (``elevator" modes). The dispersion relation for the most unstable modes with wavenumber $k=|\bk|$ is
\begin{align}
    \label{eq:D0}
    \lambda+\nu k^2+\frac{\NT^2}{\lambda+\kapT k^2}-\frac{\Nmu ^2}{\lambda+\kapmu k^2}=0,
\end{align}
where  $\lambda$ is complex. Their growth rate is the root of \Eq~\ref{eq:D0} with $\Re[\lambda]>0$. Below, we determine the growth rate, $\lambda(k)$, and its maximum value, $\lambdamax$, as a function of the parameters $\NT,\Nmu, \kapmu, \nu,$ and $\kapT$.

Without diffusion, \Eq~\ref{eq:D0} simplifies to
\begin{align}
\label{eq:D0_ideal}
    &\lambda^2+\NT^2-\Nmu ^2=0.
\end{align}
Its solutions are $\lambda=\pm i\sqrt{\NT^2-\Nmu ^2}$. These correspond to buoyancy oscillations, where perturbations from equilibrium are restored due to the thermal stratification.

Finite diffusivities can enable instability. For the typical ordering $\kapmu\ll \nu\ll\kapT$ in stellar interiors, perturbations rapidly equilibrate temperature with their surroundings while retaining their initial composition. This mitigates the restoring thermal response and allows compositionally driven motions to grow. Focusing on modes with fast thermal and slow compositional diffusion ($\kapmu k^2\ll\nu k^2,\lambda\ll\kapT k^2$), the dispersion relation (\Eq~\ref{eq:D0}) becomes
\begin{align}
    \label{eq:D0_simplified}
    \lambda^2-\Nmu ^2+\lambdad (k)\lambda&=0,\\
    \label{eq:dragrate}
    \lambdad (k)&\equiv \frac{\NT^2}{\kapT k^2}+\nu k^2,
\end{align}
where $\lambdad (k)$ is the drag rate on a perturbation due to the diffusive buoyancy response\footnote{When thermal diffusion is fast, the oscillatory motions $\lambda\sim \pm i\NT$ of perturbations in a thermally stratified (subadiabatic) fluid become overdamped $\lambda\sim- \NT^2/\kapT k^2$ and perturbations experience a drag-like force \citep{lignieres1999small,skoutnev2023critical}.} and the viscosity. 

The thermohaline growth rate is obtained by solving for the positive root of \Eq~\ref{eq:D0_simplified}. One can see that it approaches a maximum rate $\lambda(k)\approx \Nmu$ at wavenumbers $k$ where drag is negligible $\lambdad (k)\ll \Nmu $, and is otherwise reduced to $\lambda(k)\approx \Nmu^2/\lambdad(k)$ when drag is significant $\lambdad (k)\gg \Nmu $. The condition for negligible drag holds across the wavenumber range
\begin{align}
    \kT\ll k\ll \knu,
\end{align}
where 
\begin{align}
    \kT\equiv\left(\frac{\NT^2}{\kapT N_\mu}\right)^{1/2},\quad \knu\equiv\left(\frac{N_\mu}{\nu}\right)^{1/2}.
\end{align}
The wavenumber $\kT$ marks the transition $\NT^2/\kapT k^2\sim \Nmu$ where the diffusive buoyancy response becomes significant at lower $k$, with $\lambdad(k)\approx \Nmu \kT^2/k^2$  for $k\ll\kT$. Similarly, the wavenumber $\knu$ marks the transition $\nu k^2\sim \Nmu$ where viscosity becomes significant at higher $k$, with $\lambdad(k)\approx \Nmu k^2/\knu^2$ for $k\gg\knu$.

Combining the above estimates yields an approximate expression for the growth rate:
\beq
\label{eq:growthrate_summary}
  \lambda(k)\approx\Nmu\times\left\{\begin{array}{cc}
    \displaystyle{\frac{k^2}{\kT^2}} &  k\ll\kT \\
    1
    & \quad \kT\ll k\ll\knu\\
    \displaystyle{\frac{\knu^2}{k^2}} & k\gg\knu\\
           \end{array}\right. .
\eeq
This expression is valid when: (1) compositional diffusion is negligible $\kapmu k^2\ll\lambda(k)$, which requires $k\ll k_{\nu\kapmu}\equiv(\Nmu^2/\nu\kapmu)^{1/4}$ and (2) thermal diffusion is fast $\lambda(k)\ll\kappa_Tk^2$, which requires $\Nmu\ll\NT$\footnote{Thermal diffusion is not fast at $k<\kT$ when $\Nmu\gtrsim \NT/\sqrt{2}$. The analysis near Ledoux stability is presented in Appendix~\ref{ap:Ledoux}. }. The growth rate resembles a plateau with slopes on both sides (Figure~\ref{fig:GR}). It slightly peaks at the plateau's central wavenumber $\kd=\sqrt{\kT\knu}$ where the drag rate is at its minimum $\lambdad(\kd)=2\NT Pr^{1/2}$. This is the canonical ``finger" wavenumber \citep{stern1960salt}: 
\begin{align}
\label{eq:kopt}
\kd \equiv\left(\frac{\NT^2}{\nu\kapT}\right)^{1/4}.  
\end{align}

The thermohaline instability as described above is in an ``inertial" regime where compositional convection is driven by modes with a fast growth rate $\lambda\approx\lambdamax\approx \Nmu$  across a range of small scales unaffected by drag. The inertial regime requires a sufficiently large $\mu$-gradient with $\Nmu\gg \NT Pr^{1/2}$ to satisfy the condition $\kT\ll\knu$. For smaller $\mu$-gradients, drag is significant at all $k$ and the growth rate is well approximated by $\lambda(k)\approx \Nmu^2/\lambdad(k)$, which has a narrow peak at $k=\kd$ with a maximum value $\lambda_{\max}=\Nmu^2/2\NT Pr^{1/2}$. This is an inertia-free, ``drag" regime governed by slow, diffusive dynamics \citep{xie_reduced_2017,xie2019jet,fraser2025spontaneous}. Small $\mu$-gradients drive instability as long as the growth rate $\lambda_{\max}$ exceeds the compositional diffusion rate across the finger scale $\kapmu \kd^2$ (i.e., $\kd\ll k_{\nu\kapmu}$), which requires $\Nmu\gg\NT \tau^{1/2}$.

The maximum growth rate away from marginal stability $(\Nmu\gg \NT\tau^{1/2})$ can now be summarized as
\begin{align}
\label{eq:lambdamax}
    \lambdamax =\Nmu \times   
    \begin{cases}
      1 & 2Pr^{1/2}\ll\displaystyle{\frac{\Nmu }{\NT}}<1\\
      \displaystyle{\frac{\Nmu }{2\NT Pr^{1/2}}} & \tau^{1/2}\ll\displaystyle{\frac{\Nmu }{\NT}}\ll 2Pr^{1/2}
    \end{cases}     .
\end{align}
The inertial, drag, and marginal stability regimes span the instability parameter space when $\tau\ll Pr$. The inertial regime agrees with the results of \cite{brown2013chemical} obtained for $\tau\sim Pr$, while the drag regime agrees with the results of \cite{xie_reduced_2017} obtained for $\tau\ll Pr$. Scalings for $\lambdamax$ in the marginal regime (where $\Nmu\sim\NT \tau^{1/2}$) and a detailed comparison with these studies are presented in Appendix~\ref{ap:Comparison}.

Note that the instability parameter range $\tau^{1/2}<\Nmu/\NT<1$ is equivalent to the range $1<R_0<\tau^{-1}$ based on the standard definition of the density ratio
\begin{align}
    R_0=\frac{\NT^2}{\Nmu^2}.
\end{align}
The inertial regime occurs for $1<R_0\ll Pr^{-1}/4$, the drag regime occurs for $Pr^{-1}/4\ll R_0\ll \tau^{-1}$, and the marginal stability regime occurs near $R_0\sim\tau^{-1}$.

\section{Turbulent transport models}
\label{sec:TransportModels}
\begin{figure*}
    \centering
    \includegraphics[width=\linewidth]{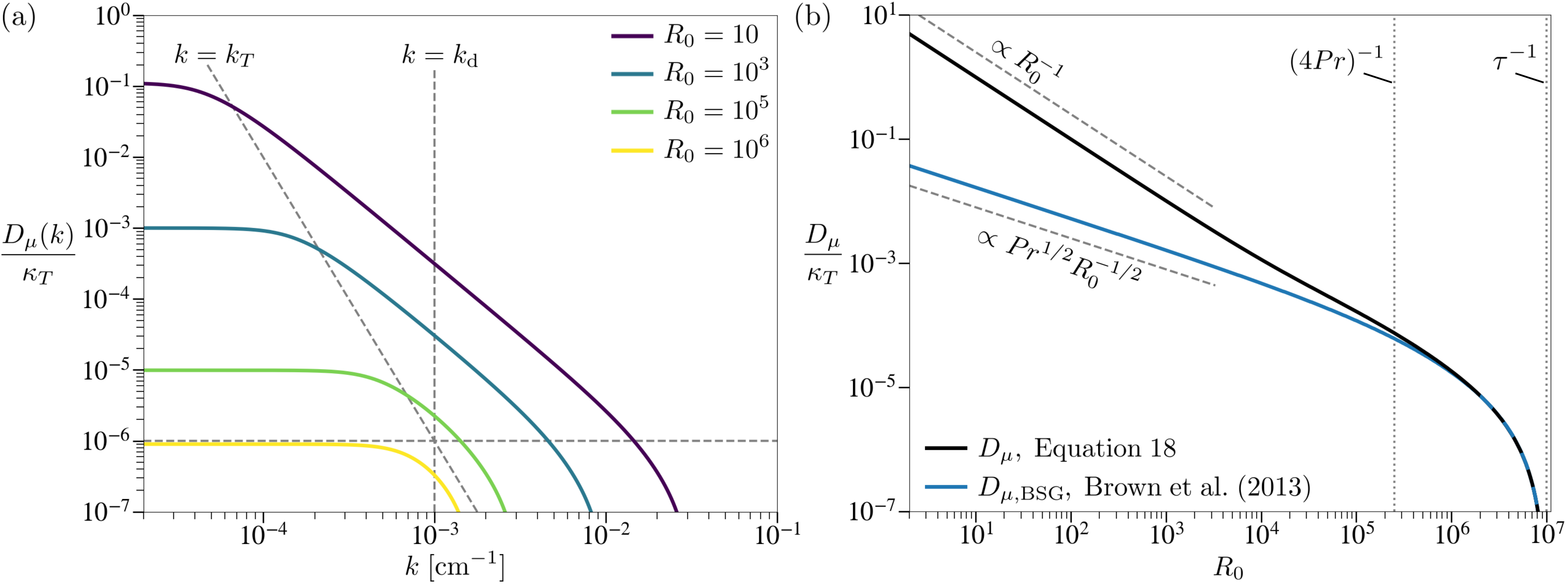}
    \caption{Turbulent diffusion coefficients  for representative parameters inside a red giant (same as panel (a) in Figure~\ref{fig:GR}). Panel (a) shows the wavenumber-dependent diffusion coefficient $D_\mu(k)=\lambda(k)/k^2$, assuming unstable modes contribute independently to vertical transport. For slow modes ($k<\kT$), $D_\mu(k)\sim \kapT /R_0$ is independent of wavenumber. For fast modes ($k>\kT$), $D_\mu(k)\sim \kapT \kT^2/R_0k^2$ decreases as $\propto 1/k^2$. Panel (b) compares two models of the total turbulent diffusion coefficient, (1) $D_{\mu }$ based on all comparably fast-growing modes (\Eq~\ref{eq:Dmu_unified} with $\alpha=e,p=4$) and (2) $D_{\mu,\rm BSG}$ based on the single fastest growing mode (Equation~\ref{eq:Brown}). We use the prefactor $C_{\rm BSG}=7$ for $D_{\mu,\rm BSG}$ (calibrated by simulations in \cite{brown2013chemical}) and we choose $C=5$ so that $D_{\mu }$ and $D_{\mu,\rm BSG}$ closely  match in the drag and marginal regimes. The models agree in the drag and marginal regimes (for $R_0>Pr^{-1}/4$) where $k\sim \kmax$ modes dominate the transport, but diverge in the inertial regime (for $R_0< Pr^{-1}/4$) where $k\sim\kT$ modes dominate  $D_{\mu }$ but $k\sim\kmax=\kd$ modes still determine $D_{\mu,\rm BSG}$.
    }
    \label{fig:Dmu}
\end{figure*}

We are interested in the turbulent transport across a fluid layer of thickness $H$ containing an inverted $\mu$-gradient. Following (multi-mode) mixing length theory \citep{stevenson1979turbulent}, steady-state convection maintains a balance between the driving of unstable modes and the nonlinear cascade. This balance is regulated by secondary (e.g. shear) instabilities \citep{brown2013chemical}. For a mode with horizontal scale $1/k$, the characteristic vertical speed $u_z(k)$ is obtained by equating the thermohaline and shear instability growth rates $\lambda(k)\sim u_z(k)k$, which also equal the nonlinear cascade rate. Fluid parcels rising/sinking over their growth timescales $\sim1/\lambda$ then transport their composition over vertical scales  $l_z\sim u_z/\lambda\sim 1/k$. Combining these relations, the scale-dependent turbulent diffusion coefficient is
\begin{align}
\label{eq:Dmu_k}
    D_\mu(k)\sim u_{z}(k) l_z\sim \frac{\lambda(k)}{k^2},
\end{align}
and the total turbulent diffusion coefficient \citep{currie2020convection} can be defined as
\begin{align}
\label{eq:Dmu_tot}
    D_\mu=\int D_\mu(k)d\ln k.
\end{align}
Turbulent transport is typically dominated by modes at either a single wavenumber or a small subset of wavenumbers \citep{barker2014theory,augustson2019model,currie2020convection}.

It is helpful to briefly review the scenario of a Ledoux-unstable fluid with $\Nmu\gg\NT$ (overturning convection). Then, modes across a wide range of wavenumbers $\kH<k<\knu$ have growth rate $\lambda\approx\Nmu$, where the lowest wavenumber $\kH\sim1/H$ is constrained by the layer thickness. The resulting diffusion coefficient is
\begin{align}
\label{eq:Dmu_H}
    D_\mu\sim\frac{\Nmu}{\kH^2} \sim\Nmu H^2\qquad (\Nmu\gg\NT).
\end{align}

In the themohaline scenario $\Nmu\ll \NT$, mixing is less efficient because thermal stratification reduces the buoyant driving of low $k$ (large scale) modes. Recall that thermal diffusion in the inertial regime enables small scales modes with $\kT<k<\knu$ to grow with the ``fast" rate $\lambda\approx \Nmu$ and large scale modes with $\kH<k<\kT$  to grow with the reduced, ``slow", rate $\lambda\approx\Nmu k^2/\kT^2$ (Section~\ref{sec:LinearTheory}). The transport may thus be multi-scale with contributions from  both types of modes. Below, we examine transport by the fast and slow modes separately, consider their possible interaction, and then propose a new model for $D_\mu$ based on \Eq~\ref{eq:Dmu_tot}.

The fast modes are the same modes with constant $\lambda(k)\approx \Nmu$ as in overturning compositional convection. Because transport-enhancing structures, such as thermohaline staircases, cannot form at sufficiently low $Pr$ \citep{traxler2011numerically,brown2013chemical,garaud2015excitation}, mixing length theory for homogeneous turbulence can be applied to thermohaline convection. Evaluating $D_\mu$ (\Eq~\ref{eq:Dmu_tot}) for fast modes in the inertial regime (when $\kT\ll\knu$), one finds that the largest scale modes with $k\sim\kT$ dominate transport\footnote{This holds away from  Ledoux stability for $2<R_0\ll Pr^{-1}/4$. Even larger scale modes may begin to dominate transport closer to Ledoux stability $R_0\rightarrow1$ (see Appendix~\ref{ap:Ledoux}).}:
\begin{align}
\label{eq:Dmu_kT}
    D_\mu\sim D_\mu(\kT)\sim\frac{\Nmu}{\kT^2} \sim\frac{\kapT}{R_0}.
\end{align}
Interestingly, this is the scaling in the traditional Kippenhahn model. We note that \cite{brown2013chemical} proposed a theory for $D_\mu$ based on the fastest growing modes with growth rate $\lambdamax$ and wavenumber $\kmax$:
\begin{align}
    \label{eq:Brown}
    D_{\mu,\rm BSG}=\frac{C_{\rm BSG}^2\lambdamax^2}{\kmax^2(\lambdamax+\kappa_\mu \kmax^2)},
\end{align}
where $C_{\rm BSG}$ is a constant and the new factor in the denominator accounts for the reduction of turbulent transport by compositional diffusion (which is significant near marginal stability). In the inertial regime (far from marginal stability), the fastest modes have $\kmax= \kd$, $\lambdamax=\Nmu\gg\kappa_\mu \kd^2$, and their transport is
\begin{align}
\label{eq:Dmu_kd}
     D_{\mu,\rm BSG}\sim D_\mu(\kd)\sim \frac{\Nmu}{\kd^2}=\frac{\kapT Pr^{1/2}}{R_0^{1/2}}.
\end{align}
This is smaller than the transport from $\kT$ modes (\Eq~\ref{eq:Dmu_kT}) by a factor of $(\kT/\kd)^2=(R_0Pr)^{1/2}$, which becomes significant deep inside the inertial regime $R_0\ll Pr^{-1}$. The difference between $D_{\mu}(\kT)$ and $D_{\mu}(\kd)$ can be seen in Figure~\ref{fig:Dmu} where $D_\mu(k)$ is numerically evaluated for typical parameters in a red giant star. 

We now turn to the slow modes with $k<\kT$, whose transport was examined in \cite{ulrich1972thermohaline} and \cite{kippenhahn1980time}. They identify the linear diffusive buoyancy dynamics (where growth is rate limited by thermal diffusion) and then argue on dimensional grounds that $D_\mu(k)\sim u_z(k)/k\sim \lambda(k)/k^2$. The quadratic dependence of $\lambda=k^2(\Nmu /\kT^2)$ on wavenumber gives the traditional estimate $D_\mu(k)\sim \Nmu/\kT^2=\kapT/R_0$, where all large scales contribute equally. This matches the $D_\mu(k)$ from the fast modes  (\Eq~\ref{eq:Dmu_kT}) at $k=\kT$ where the crossover of $\lambda(k)$ between the two mode types occurs (also see Figure~\ref{fig:Dmu}). \Eq~\ref{eq:Dmu_tot} implies that the total transport from the slow modes follows
\begin{align}
    D_\mu\approx\frac{\kapT}{R_0}\ln\left(\frac{\kT}{\kH}\right),
 \end{align}
which is  larger than that of the fast modes (\Eq~\ref{eq:Dmu_kT}) by a logarithmic factor. However, we caution that the nonlinear saturation of slow modes has yet to be examined in simulations. It is unclear how the diffusive buoyancy dynamics at large scales affects the secondary instabilities and whether \Eq~\ref{eq:Dmu_k} applies. 

The estimates above neglect the possible nonlinear interactions between the simultaneously driven slow and fast modes. We argue that the slow, large scale modes may be suppressed by fast horizontal mixing from the small scale modes. The slow modes require a compositional perturbation with a large horizontal scale $1/k$ to remain coherent while they grow. However, these horizontal gradients are erased by horizontal turbulent diffusion from the saturated fast modes with $k\sim\kT$  at the same rate  $ D_{\mu,h}(\kT)k^2\sim \Nmu k^2/\kT^2$ as the slow mode grow rate $ \lambda(k)$. Here, $D_{\mu,h}(k)\sim u_h(k)/k$ is estimated assuming near-equipartition $u_h(k)\sim u_z(k)$. As a result, the slow modes may be suppressed and only the fast modes then provide transport.

Combining the above results, a modified expression for $D_\mu$ valid in all thermohaline regimes can be designed with two changes to \Eq~\ref{eq:Dmu_tot}: (1) including the compositional diffusion reduction factor in the denominator (see \Eq~\ref{eq:Brown}) and (2) using adaptive integration limits that shift with the regime. In the inertial regime, the integration limits should cover the fast modes: $\kT\leq k\leq \knu$. In the drag and marginal regimes, the limits should instead cover a window around the optimal wavenumber, $\kmax$, in order to match $D_{\mu,\rm BSG}$: $\kmax/\alpha<k<\alpha \kmax$ with $\alpha\gtrsim1$. We define lower and upper limits that smoothly interpolate between regimes: $k_1=||(k_T,\kmax/\alpha)||_{-p}$ and $k_2=||(k_\nu,\alpha \kmax)||_{p}$, where $||(a,b)||_p=(|a|^p+|b|^p)^{1/p}$. For large $p>1$, the $p$-norms behave like smooth minimum and maximum functions for $k_1$ and $k_2$, respectively. With these changes,  \Eq~\ref{eq:Dmu_tot} becomes
\begin{align}
    \label{eq:Dmu_unified}
    D_{\mu}&=\int_{k_{1}}^{k_{2}} \frac{C^2\lambda(k)^2}{k^2(\lambda(k)+\kapmu k^2)}\Theta(\lambda(k))d \ln k,
\end{align}
where $C$ is analogous to $C_{\rm BSG}$, and the Heaviside function $\Theta$ ensures that only unstable modes contribute. Comparison of  $D_{\mu}$ against $D_{\mu,\rm BSG}$  (Figure~\ref{fig:Dmu}) confirms their agreement in the drag and marginal regimes and divergence in the inertial regime.

\section{Discussion}
\label{sec:Discussion}

Linear theory reveals that thermohaline convection is driven across a wide range of scales in the limit of $Pr\rightarrow 0$ and large inverted $\mu-$gradients $2<R_0\ll Pr^{-1}/4$: at large scales by slow modes (for $k_H< k<\kT$) and at small scales by fast modes (for $\kT< k<\knu$). Turbulent diffusion arguments predict that the modes with $k\sim \kT$ self-consistently control the vertical transport because (1) they are the largest scale fast mode and (2) their fast horizontal mixing suppresses the competing slow modes. The $\kT$ modes provide a turbulent diffusivity $D_\mu\sim \kapT/R_0$  equivalent to the traditional scaling proposed in \cite{ulrich1972thermohaline,kippenhahn1980time}.

Testing these predictions with simulations poses several challenges. Current numerical models typically use 3D domains large enough to fit multiple wavelengths of the marginally fastest growing modes,  $2\pi/\kd$ \citep{garaud20152d}. However, capturing transport by all the fast modes necessitates larger domains, several times $2\pi/\kT$. Ensuring scale separation between $\kT$ and $\kd$ modes requires reaching the challenging limit $\kT/\kd=(R_0Pr)^{1/4}\ll1$ where the Reynolds number is large, $Re\sim\Nmu/\nu \kT^2\sim 1/R_0Pr$. As an example, panel (b) of Figure~\ref{fig:GR} shows how their scale separation widens as $Pr$ is decreased from previously numerically accessible values ${\cal O}(10^{-2})$ toward stellar values ${\cal O}(10^{-6})$ at fixed $R_0=10,\tau=10^{-7}$. Future simulations pushing to orders of magnitude lower $Pr$ may begin to probe whether transport dominance shifts from $k\sim\kd$ modes to $k\sim\kT$ modes, which can be disentangled using spectral diagnostics. Capturing slow modes is even more challenging, requiring (1) box sizes $L\gg 2\pi/\kT$ and (2) simulation run times long enough to diagnose their growth or suppression. Future numerical efforts may benefit from alternative nondimensionalizations and reduced models \citep{lignieres1999small,xie_reduced_2017,cope2020dynamics,fraser2025spontaneous,fraser2026bridging}.

We anticipate that a complete theory for $D_\mu(R_0,Pr,\tau)$ at low $Pr$ will exhibit smooth transitions across stability or regime-change boundaries in the parameter range $0<R_0<\tau^{-1}$. When $R_0$ increases from $R_0=0$ across the Ledoux stability boundary $R_0=1$, $D_\mu$ should transition from the viscosity-free scaling for overturning convection at $R_0\ll1$ (\Eq~\ref{eq:Dmu_H}) to another viscosity-free scaling at $R_0>1$ (e.g., see \cite{castro2024fast}), which $D_\mu\sim \kapT/R_0\propto\nu^{0}$ satisfies\footnote{Thermohaline instability is enabled by thermal diffusion and is most active when (1) $\mu$-gradients are large and (2) viscosity is negligible. The  scaling $D_\mu\sim \kapT Pr^{1/2}/R_0^{1/2}\propto \nu^{1/2}$ \citep{brown2013chemical} seems unlikely to hold near Ledoux stability $R_0\sim1$  since it vanishes in the inviscid limit $Pr\rightarrow0$.}. The details of this transition are presented in Appendix~\ref{ap:Ledoux}, and it occurs over the range $0<R_0<2$. Further increasing $R_0$ toward the boundary between the inertial and drag regimes, $R_0= Pr^{-1}/4$, the viscosity-free scaling $D_\mu\sim \kapT/R_0$ cannot continue to hold because viscous effects become significant at the driving scales. The viscosity-dependent scaling $D_\mu\sim \kapT Pr^{1/2}/R_0^{1/2}$ should take over \citep{brown2013chemical}. Thus, it is plausible that $D_\mu$ shifts from
\beq
\label{eq:D_mu_final1}
  D_\mu\sim\displaystyle{\frac{\kapT}{R_0}}\quad \mathrm{near}\quad  R_0\gtrsim2
\eeq
to
\beq
\label{eq:D_mu_final2}
  D_\mu\sim \kapT\displaystyle{\frac{Pr^{1/2}}{R_0^{1/2}}} \quad\mathrm{near}  \quad R_0\lesssim Pr^{-1}/4
\eeq
as transport dominance shifts from the fast modes with $k\sim \kT$ to those with $k\sim \kd$\footnote{The increase of $D_\mu$ at low $R_0$ resurfaces the possibility of layer formation by the collective instability, which favors larger $D_\mu/\kapT$. \cite{garaud2015excitation} found that the collective instability is suppressed below $Pr={\cal O}(10^{-3})$ based on $D_\mu/\kapT\sim  Pr^{1/2}/R_0^{1/2}$. More efficient mixing $D_\mu/\kapT\sim  1/R_0$ would lower the suppression threshold.}. Finally, the scaling of $D_\mu$ in the drag regime, $Pr^{-1}/4<R_0<\tau^{-1}$, depends on the viscosity \citep{xie_reduced_2017,xie2019jet,fraser2025spontaneous}. An expression for $D_\mu$ naturally capturing these transitions in the thermohaline regimes ($R_0>2$) is proposed in \Eq~\ref{eq:Dmu_unified}.

The question of whether transport is dominated by the fast mode with the largest scale or the marginally fastest growth rate has also been raised for rotating convection. That system (e.g. for aligned rotation and gravity with $\Omega\gg|N_T|$) similarly exhibits a spectrum of ``fast" modes with $\lambda\sim|N_T|$ for wavenumbers where rotation and viscosity are negligible. The spectrum has a lowest wavenumber $\sim k_H\Omega/|N_T|$ and a marginally fastest growing wavenumber $\sim k_H(\Omega H^2/\nu)^{1/3}$ that are analogous to $\kT$ and $\kd$. Simulations of internally heated rotating convection  (see \cite{barker2014theory,currie2020convection} and references within) have shown that the lowest wavenumbers dominate transport, supporting the mixing length theory of \cite{stevenson1979turbulent}. Confirming analogous transport dominance by $\kT$ modes in simulations of thermohaline convection would provide support for a universal theory of convective transport. Otherwise, nonlinear suppression of $\kT$ modes in the thermohaline case would need to be explained.

The ``ultimate" regime of thermohaline convection (the limit of $Pr\rightarrow0$ and $R_0\ll Pr^{-1}$)  is less applicable when $Pr$ is only moderately small (e.g. interiors of crystallizing white dwarfs where $Pr = {\cal O}(10^{-2})$ \citep{montgomery2024fluid}) or when thermohaline convection is only marginally excited (e.g. in the planetary-engulfment context \citep{theado2012metal,behmard2023planet}). Then, the relative growth of $k\sim \kd$ modes is sufficiently fast that they dominate transport and $D_\mu$ follows the model in \cite{brown2013chemical}.

\begin{acknowledgments}
We thank Adrian Fraser, Adrian Barker, Andrei Beloborodov, Jim Fuller, and Yuri Levin for helpful discussions and Adrian Fraser for significant feedback during the development of these ideas. 
\end{acknowledgments}

\appendix

\section{Comparison with previous work and asymptotics near marginal stability}
\label{ap:Comparison}

The maximum growth rate $\lambdamax$ in the inertial and drag regimes was derived in Section~\ref{sec:LinearTheory}. Here we compare our results with those of \cite{brown2013chemical} and \cite{xie_reduced_2017} and provide scalings for $\lambdamax$ near marginal stability.

\cite{brown2013chemical} consider the limit $\tau\sim Pr$ and categorize the instability parameter space into three regimes based on the reduced density ratio
\beq
    r\equiv\frac{R_0-1}{\tau^{-1}-1}, \quad 0<r<1,
\eeq
where $r=0$ is Ledoux stability ($R_0=1$) and  $r=1$ is marginal stability ($R_0=\tau^{-1}$). Their first two regimes are away from marginal stability at $0<r\ll Pr$ and $Pr\ll r\ll1$, and their third regime is near marginal stability at $Pr\ll r\lesssim1$. The dimensional form of $\lambdamax$ in the first and second regimes (\Eqs~B5 and B15 in \cite{brown2013chemical})  is reproduced below:
\beq
\label{eq:BSG13growthrate_summary}
  \lambdamax\approx
    \left\{\begin{array}{cc}
    N_T &  1<R_0\ll1+Pr/\tau \\
    \Nmu & 1+Pr/\tau\ll R_0\ll \tau^{-1}\\
           \end{array}\right. ,
\eeq
where we have used the approximation $r\approx\tau(R_0-1)$ appropriate for the limit $\tau\ll1$. These are both consistent with the inertial regime defined in Section~\ref{sec:LinearTheory}  (where $\lambdamax\approx\Nmu$  for $1<R_0\ll Pr^{-1}/4$) since $N_T=R_0^{1/2}\Nmu\approx \Nmu$ in the first regime of \Eq~\ref{eq:BSG13growthrate_summary}  and  $\tau^{-1}\sim Pr^{-1}$ is assumed in the second regime.

The third regime in \cite{brown2013chemical} examined the transition from the inertial to the marginal stability regime where the growth rate vanishes $\lambdamax\rightarrow 0$ as the density ratio increases to $R_0\rightarrow\tau^{-1}$. This is appropriate in the case $\tau\sim Pr$ considered in \cite{brown2013chemical} since the drag regime $Pr^{-1}/4\ll R_0\ll\tau^{-1}$ is absent. However, when $\tau\ll Pr$, the growth rate first decreases as $\lambdamax\approx\Nmu/(4 R_0 Pr)^{1/2}$ with increasing $R_0$ in the drag regime (Section~\ref{sec:LinearTheory}) before transitioning to the marginal regime. This behavior was analyzed in \cite{xie_reduced_2017} using their ``IFSC" model, which drops the inertia term from the momentum equation. For completeness, we present below the transition from the drag to the marginal regime using the approach in \cite{xie_reduced_2017}. 

\begin{figure}
    \centering
    \includegraphics[width=1.0\linewidth]{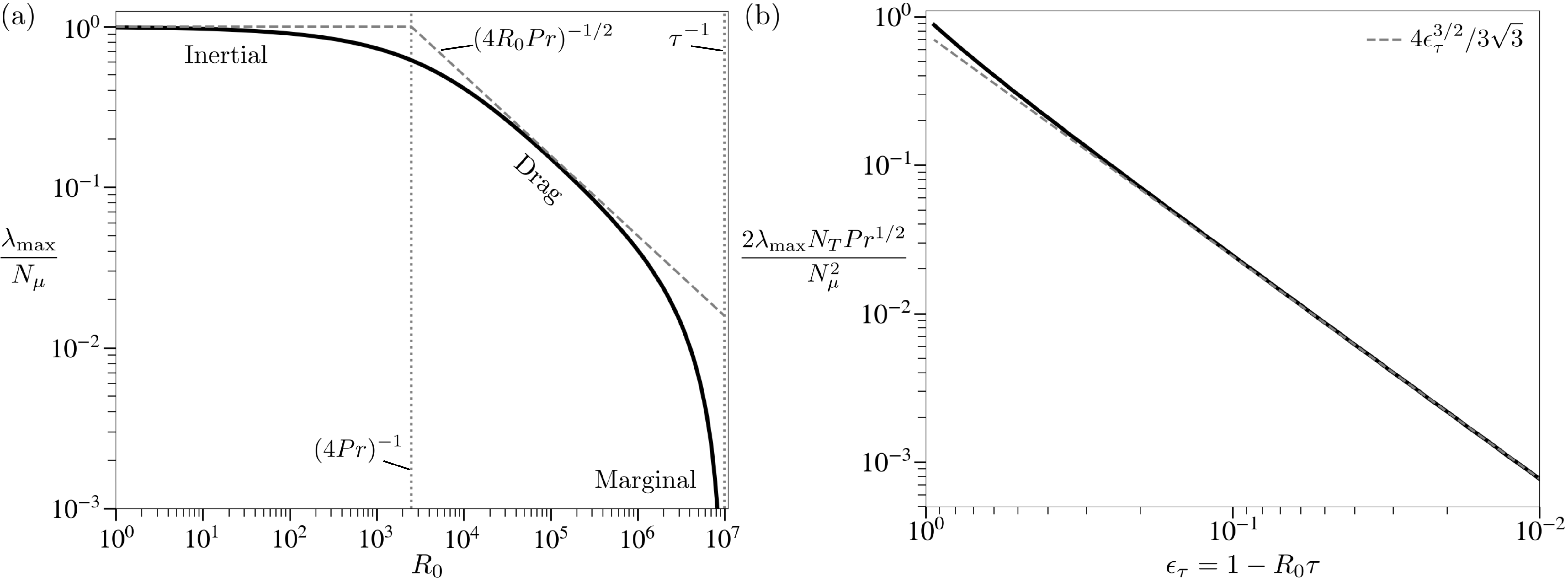}
    \caption{  Maximum growth rate $\lambdamax$ versus density ratio $R_0=\NT^2/\Nmu^2$ for illustrative parameters $Pr=10^{-4}$, $\tau=10^{-7}$. Numerical solutions (black curves) of the full dispersion relation (\Eq~\ref{eq:D0}) for $\lambdamax$ are compared against approximate analytic expressions developed in the text (dashed gray curves). Panel (a) compares $\lambdamax$ in the inertial and drag regimes against the approximations in \Eq~\ref{eq:lambdamax}. Panel (b) compares  $\lambdamax$ in the marginal regime against the first approximation in \Eq~\ref{eq:lambdamarg} for the limit $\epsilon_\tau\ll1$.
    }
    \label{fig:Lambdamax}
\end{figure}

To analyze the  dispersion relation near marginal stability, we define the small parameter $\epsilon_\tau=1-R_0\tau$. The drag and marginal regimes differ only in that compositional diffusion becomes significant near marginal stability; the drag and thermal diffusion rates remain fast, $\lambda(k) \ll\lambdad(k),\kapT k^2$. Under these conditions, \Eq~\ref{eq:D0} becomes
\begin{align}
    \label{eq:Dmarg}
    \lambdad(k)-\frac{\Nmu ^2}{\lambda+\kapmu k^2}=0.
\end{align}
The solution for the dimensional growth rate is $\lambda(k)=\Nmu^2/\lambdad(k)-\kappa_\mu k^2$. It can be made nondimensional using the scales in the drag regime $\lambda=\lambda'N_\mu^2/2 N_TPr^{1/2}$ and $k=k'\kd$ (primed quantities are nondimensional), yielding
\begin{align}
    \label{eq:lambdamarg_k}
    \lambda'(k')=\frac{2 k'^2}{1+k'^4}-2(1-\epsilon_\tau)k'^2.
\end{align}
Optimizing this expression over $k'$ gives the maximum growth rate. The optimal wavenumber $\kmax$ is
\begin{align}
    \left(\frac{\kmax}{\kd}\right)^4=\frac{\sqrt{9-8\epsilon_\tau}-3+2\epsilon_\tau}{2(1-\epsilon_\tau)}\approx\left\{\begin{array}{cc}
    \displaystyle{\frac{\epsilon_\tau}{3}} &  \epsilon_\tau\ll1 \\
    1-4(1-\epsilon_\tau) & \epsilon_\tau\rightarrow 1\\
           \end{array}\right. .
\end{align}
Away from marginal stability ($ \epsilon_\tau\rightarrow 1$), the optimal wavenumber approaches the finger wavenumber $\kmax\rightarrow \kd$ that minimizes the drag rate, but near marginal stability ($ \epsilon_\tau\rightarrow 0$) it decreases $\kmax\propto \kd \epsilon_\tau^{1/4}$ to mitigate suppression by compositional diffusion. The maximum growth rate $\lambdamax=\lambda(\kmax)$ is  
\beq
\label{eq:lambdamarg}
  \lambdamax=\frac{\Nmu^2}{2N_TPr^{1/2}}\frac{\sqrt{2}\left(3-\sqrt{9-8\epsilon_\tau}\right)\sqrt{(1-\epsilon_\tau)(\sqrt{9-8\epsilon_\tau}-3+2\epsilon_\tau})}{\sqrt{9-8\epsilon_\tau}-1}\approx\displaystyle{\frac{\Nmu^2}{2N_TPr^{1/2}}}\left\{\begin{array}{cc}
    \displaystyle{\frac{4\epsilon_\tau^{3/2}}{3\sqrt{3}}} &  \epsilon_\tau\ll1 \\
    1 -2(1-\epsilon_\tau) & \epsilon_\tau\rightarrow 1\\
           \end{array}\right.,
\eeq
which transitions between the drag and marginal regimes at $\epsilon_\tau\sim1$. These results agree with \cite{xie_reduced_2017} (their \Eqs~21, 24, 25, and 26). Comparison of the numerical solution and the asymptotic expressions for the inertial, drag, and marginal regimes (\Eqs~\ref{eq:lambdamax} and~\ref{eq:lambdamarg}) is presented in Figure~\ref{fig:Lambdamax}.

\section{Asymptotics near Ledoux stability}
\label{ap:Ledoux}

This section analyzes the growth rate $\lambda(k)$ and  turbulent diffusion coefficient $D_\mu(k)$ around the Ledoux stability boundary $R_0=1$. We focus on the affected low wavenumber modes $k<k_T$ where thermal diffusion transitions from being fast to slow (i.e. $\kapT k_T^2/\lambda(k_T)\approx R_0\sim1$) between the limits of thermohaline $R_0\gg1$ and overturning (Ledoux unstable) $R_0\rightarrow0$  convection.

Near Ledoux stability, diffusion of momentum and composition is negligible for the low wavenumber modes of interest $k<\kT$ (i.e., $\nu k^2,\kapmu k^2\ll \lambda$), and the dispersion relation in \Eq~\ref{eq:D0} simplifies to
\begin{align}
    \label{eq:D_ledoux_1}
    \lambda+\frac{\NT^2}{\lambda+\kapT k^2}-\frac{\Nmu ^2}{\lambda}=0.
\end{align}
It is convenient to define a parameter $\epsilon_1\equiv R_0-1$ that is small near Ledoux stability and nondimensionalize this equation using scales in the inertial regime $\tilde \lambda =\lambda/\Nmu$, $\tilde k=k/\kT$, yielding
\begin{align}
    \label{eq:D_ledoux_2}
    \frac{\tilde\lambda^3}{1+\epsilon_1}+\tilde k^2\tilde \lambda^2+\frac{\epsilon_1}{1+\epsilon_1}\tilde \lambda-\tilde k^2=0.
\end{align}
We use dominant balance arguments below to derive asymptotic expressions for the growth rate (positive root of \Eq~\ref{eq:D_ledoux_2}) in the thermohaline limit $\epsilon_1\gg1$, near Ledoux stability $\epsilon_1\sim0$, and in the overturning limit $\epsilon_1\sim-1$.

In the thermohaline limit $\epsilon_1\gg 1$ ($R_0\gg 2$), the growth rate at low wavenumbers $\tilde k\ll1$ comes from a dominant balance between the linear term in $\tilde \lambda$ and the constant term $-\tilde k^2$. At  $\tilde k\gg1$, the balance switches to the quadratic term in $\tilde \lambda$ and the constant term $-\tilde k^2$. One can trace back from \Eq~\ref{eq:D_ledoux_1} that the cubic term remaining small at all $\tilde k$ is equivalent to thermal diffusion remaining fast. The approximate expression for $\tilde \lambda(\tilde k)$ is 
\beq
  \epsilon_1\gg 1:\quad\tilde \lambda(\tilde k)\approx\left\{\begin{array}{cc}
    \displaystyle{\frac{1+\epsilon_1}{\epsilon_1}}\tilde k^2  & \tilde k\ll1 \\
    1 &  \tilde k\gg1
           \end{array}\right. ,
\eeq
which agrees with the dimensional expressions in \Eq~\ref{eq:growthrate_summary}  since $(1+\epsilon_1)/\epsilon_1\approx1$ in the limit  $\epsilon_1\gg1$.

Near Ledoux stability $|\epsilon_1|\ll1$, the dispersion relation (\Eq~\ref{eq:D_ledoux_2}) reduces to
\begin{align}
    \label{eq:D_ledoux_3}
    \tilde\lambda^3+\tilde k^2\tilde \lambda^2+\epsilon_1\tilde \lambda-\tilde k^2=0.
\end{align}
The growth rate in \Eq~\ref{eq:D_ledoux_3} differs from the thermohaline limit at $\tilde k\lesssim1$ (it is unchanged for $\tilde k\gg1$) because the cubic term becomes significant (thermal diffusion is no longer fast). The cubic term $\tilde \lambda^3$ and constant term $-\tilde k^2$ must instead balance at $\tilde k\lesssim1$, resulting in $\tilde \lambda\approx \tilde k^{2/3}$ for $\tilde k\ll1$. However, this balance cannot continue to hold at sufficiently low $\tilde k$ because the linear term $\epsilon_1\tilde \lambda$ becomes significant at $\tilde k\sim |\epsilon_1|^{3/4}$. When the fluid is marginally Ledoux stable $\epsilon_1>0$,  the dominant balance at smaller wavenumbers changes back to the linear term $\epsilon_1\tilde \lambda$ and constant term $-\tilde k^2$, resulting in $\tilde \lambda\approx  \tilde k^{2}/\epsilon_1$ for $\tilde k\ll \epsilon_1^{3/4}$. The growth rate for $0<\epsilon_1\ll1$ has two power laws at low $\tilde k$:
\beq
\label{eq:eps1_gtr}
  0<\epsilon_1\ll1:\quad\tilde \lambda(\tilde k)\approx\left\{\begin{array}{cc}
      \displaystyle{\frac{\tilde k^{2}}{\epsilon_1}} & \tilde k\ll\epsilon_1^{3/4} \\
    \tilde k^{2/3} & \epsilon_1^{3/4}\ll\tilde k\ll 1 \\
    1 &  \tilde k\gg1 \\
           \end{array}\right. .
\eeq
\begin{figure}
    \centering
    \includegraphics[width=0.6\linewidth]{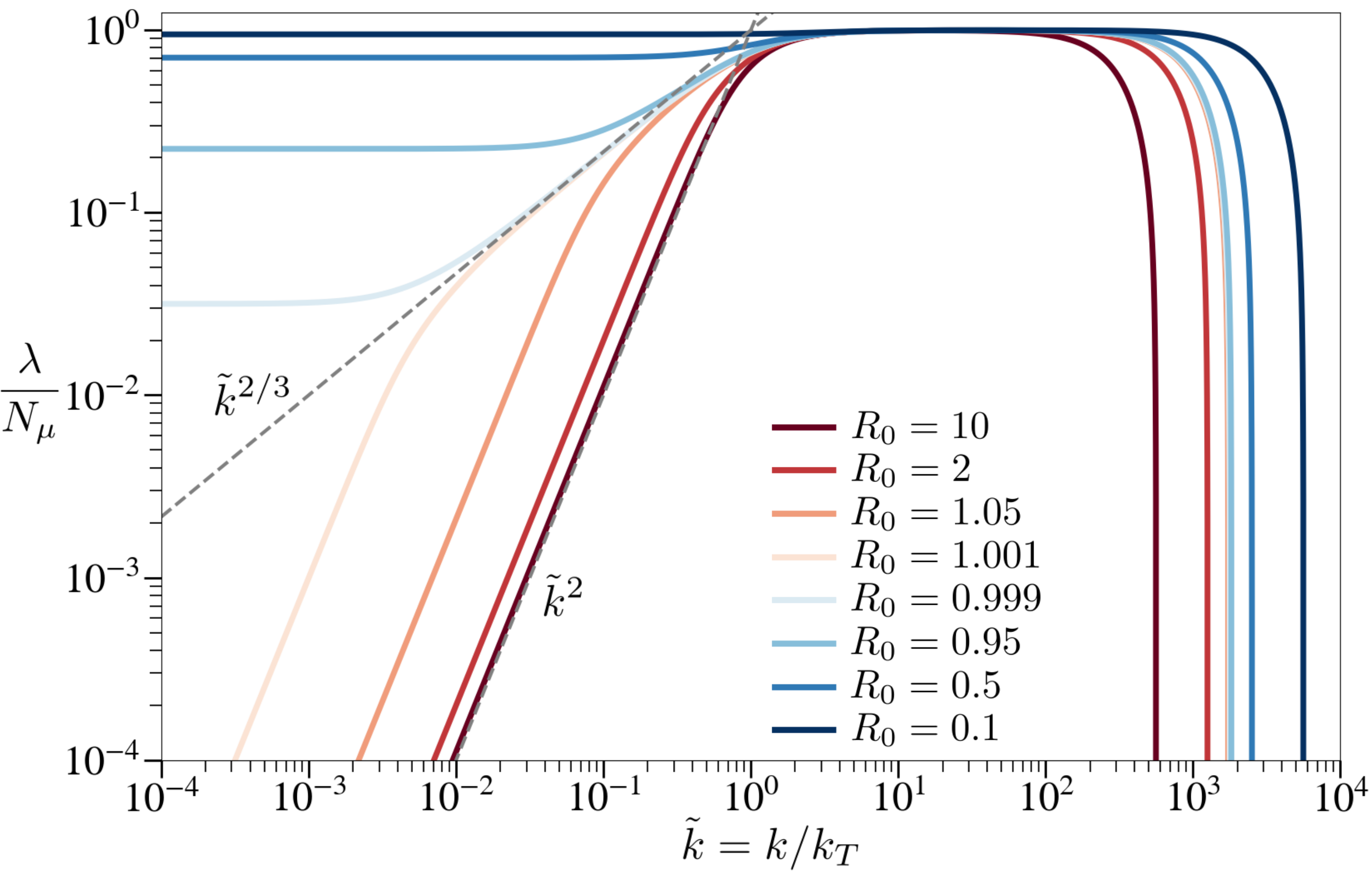}
    \caption{ Thermohaline growth rate $\lambda(k)$ versus wavenumber as the density ratio $R_0=N_T^2/\Nmu^2$ is varied across the Ledoux stability boundary $R_0=1$.
    The growth rate at low wavenumbers ($k<k_T$) smoothly increases with decreasing $R_0$ and becomes nearly flat in the overturning convection limit $R_0\ll1$.
    The two scalings $\lambda\propto k^{2/3}$ and $\lambda\propto k^{2}$ found in \Eqs~\ref{eq:eps1_gtr} and~\ref{eq:eps1_less} (valid in the transition regime $R_0\sim 1$ at $k<k_T$) are shown for reference as dashed gray lines.
    }
    \label{fig:Ledoux}
\end{figure}
At exactly $\epsilon_1=0$, the growth rate at $\tilde k<1$ (\Eq~\ref{eq:eps1_gtr}) has only the single power law $\tilde \lambda=\tilde k^{2/3}$. When the fluid is marginally Ledoux unstable $-1\ll\epsilon_1<0$, the dominant balance at lower wavenumbers instead changes to the cubic and linear terms in $\tilde \lambda$, resulting in $\tilde \lambda\approx |\epsilon_1|^{1/2}$ for $\tilde k\ll|\epsilon_1|^{3/4}$. The growth rate for $-1\ll\epsilon_1<0$ has two flat regions connected by a power law:
\beq
\label{eq:eps1_less}
  -1\ll\epsilon_1<0:\quad\tilde \lambda(\tilde k)\approx\left\{\begin{array}{cc}    
  \displaystyle{|\epsilon_1|^{1/2}} & \tilde k\ll|\epsilon_1|^{3/4} \\
    \tilde k^{2/3} & |\epsilon_1|^{3/4}\ll\tilde k\ll 1 \\
    1 &  \tilde k\gg1 
           \end{array}\right. .
\eeq
In the overturning limit $\epsilon_1\rightarrow-1$, the  power law region disappears and the two flat regions merge with $\tilde \lambda\approx 1$. Comparison of these asymptotics against the full numerical solution is shown in Figure~\ref{fig:Ledoux}, which also helps visualize how $\tilde \lambda (\tilde k)$ smoothly transitions from a plateau profile in the thermohaline limit to a flat profile in the overturning limit.

The above expressions for the growth rate can be used to evaluate $D_\mu$ across the parameter range $0<R_0<2$. The resulting scalings should connect the diffusivity-dependent scaling $D_\mu\sim \kapT/R_0$ in the thermohaline limit  $R_0\gtrsim2$ and the diffusivity-free scaling $D_\mu\sim \Nmu/k_H^2$ in the overturning limit $R_0\rightarrow0$. We assume that transport is dominated by a single mode and evaluate $D_\mu\sim \max_k \lambda(k)/k^2$ at each $R_0$ under the constraint $k\geq k_H$. This takes into account the finite layer height, otherwise $D_\mu$ would diverge for $R_0\leq 1$. Using \Eqs~\ref{eq:eps1_gtr} and~\ref{eq:eps1_less}, we find
\beq
  D_\mu(R_0)\sim \left\{\begin{array}{cc}
    \displaystyle{\frac{\Nmu\left(1-R_0\right)^{1/2}}{k_H^2}} &  0<R_0\ll 1-\left(\frac{k_H}{k_T}\right)^{4/3} \\
    \displaystyle{\frac{\Nmu}{k_H^2}}\left(\frac{k_H}{k_T}\right)^{2/3} & 1-\left(\frac{k_H}{k_T}\right)^{4/3}\ll R_0\ll 1+\left(\frac{k_H}{k_T}\right)^{4/3} \\
    \displaystyle{\frac{\kappa_T}{R_0(R_0-1)}} & 1+\left(\frac{k_H}{k_T}\right)^{4/3}\ll R_0\ll 2 \\
    \displaystyle{\frac{\kappa_T}{R_0}} &   R_0\gtrsim 2\\
    \end{array}\right. ,
\eeq
where $k_T\approx\sqrt{N_T/\kapT}$ near $R_0\sim1$ and the layer height is assumed to be large $k_H\ll\kT$. The dominant transport wavenumber is $k\sim \kT$ for $R_0\gtrsim2$ and begins to decrease as $k\sim k_T(R_0-1)^{3/4}$ for $R_0\lesssim2$, resulting in a  super-Kippenhahn scaling for $D_\mu$. The dominant transport wavenumber eventually reaches the minimum wavenumber $k\sim k_H$ near $R_0=1+(k_H/k_T)^{4/3}$ and then remains at $k_H$ at lower $R_0$ while $\lambda(k_H)$ increases following  \Eq~\ref{eq:eps1_less}. The result is $D_\mu$ scaling as $\sim \Nmu/k_H^2$ in the overturning limit $R_0\rightarrow0$.


\vspace{5mm}
% \facilities{HST(STIS), Swift(XRT and UVOT), AAVSO, CTIO:1.3m,


% \software{astropy \citep{2013A&A...558A..33A,2018AJ....156..123A},  
%           Cloudy \citep{2013RMxAA..49..137F}, 
%           Source Extractor \citep{1996A&AS..117..393B}
%           }




\bibliography{refs}{}
\bibliographystyle{aasjournal}



\end{document}